\documentclass[12pt]{article}
\usepackage{amssymb}
\begin{document}

\def\kbar{{\mathchar'26\mkern-9muk}}  
\def\vev#1{\langle #1 \rangle}
\def\tr{\mbox{Tr}}
\def\ad{\mbox{ad}\,}
\def\ker{\mbox{Ker}\,}
\def\m@th{\mathsurround=0pt}
\def\eqalign#1{\null\,\vcenter{\openup 3pt \m@th
\ialign{\strut\hfil$\displaystyle{##}$&$\displaystyle{{}##}$\hfil
\crcr#1\crcr}}\,}

\title{On Poisson Structure and Curvature}

\author{J. Madore \\
        Laboratoire de Physique Th\'eorique et Hautes 
        Energies\thanks{Laboratoire associ\'e au CNRS, \mbox{URA D0063}} \\ 
        Universit\'e de Paris-Sud, B\^at. 211, F-91405 Orsay \\
       }

\date{June, 1997}

\maketitle

\abstract{We consider a curved space-time whose algebra of functions is 
the commutative limit of a noncommutative algebra and which has
therefore an induced Poisson structure. In a simple example we determine
a relation between this structure and the Riemann tensor.}

\vfill
\noindent
LPTHE Orsay 97/25
\medskip
\eject

\parskip 4pt plus2pt minus2pt

\section{Motivation and notation}

One of the more interesting problems of noncommutative geometry (in
physics) is the study of the commutative (`classical') limit. In a
previous article (Dubois-Violette {\it et al.} 1997) we considered the
case where the commutative limit was Poincar\'e invariant and we analyzed
in detail the Poisson structure which necessarily remains as a `shadow'.
Here we suppose that the limit is a curved manifold and we discuss the
relation between the Poisson structure and the curvature.

Let ${\cal A} = {\cal C}(V)$ be the algebra of (smooth, complex-valued)
functions on a (real) space-time $V$, which for simplicity we shall
suppose diffeomorphic to ${\mathbb R}^4$, and let $\Omega^*({\cal A})$ be
the algebra of de~Rham differential forms. The space 
$\Omega^1({\cal A})$ of 1-forms is free of rank 4 as an 
${\cal A}$-module; there exists a globally defined moving frame
$\theta^\alpha$.  It can be argued that space-time should be more
properly described by a noncommutative $*$-algebra ${\cal A}_\kbar$ over
the complex numbers with four hermitian generators $q^\lambda$.  The
parameter $\kbar$ is a fundamental area scale which is presumably of the
order of the Planck area $G\hbar$.  We refer, for example, to Madore
(1997) for a synopsis of the arguments.  Let ${\cal A}_0$ be the
commutative limit of ${\cal A}_\kbar$. We shall suppose that the algebra
${\cal A}_\kbar$ is a deformation of ${\cal A}_0$ in the sense of
Gerstenhaber (1964).  The relation between ${\cal A}_\kbar$ and the
classical space-time $V$ is given (Dubois-Violette {\it et al.} 1997) by
the inclusion relation ${\cal A} \subset {\cal A}_0$.  We shall assume
that we can identify ${\cal A}_0 = {\cal C}(V_0)$ as the algebra of
(smooth, complex-valued) functions on a (real) parallelizable manifold
$V_0$ and that there is a projection of $V_0$ onto $V$.

One can consider $V_0$ as a Kaluza-Klein extension of $V$ by a manifold
which is perhaps of infinite dimension and in general not compact. It
should be stressed however that $V_0$ is a mathematical fiction. The
`real' world is described by the algebra ${\cal A}_\kbar$; it is this
algebra which we consider to be the correct Kaluza-Klein extension of
$V$ (Madore 1989b, Madore \& Mourad 1996a).  The difference in dimension
between $V_0$ and $V$ is one of the measures of the extent to which the
verb `to quantize' as applied to the coordinates of $V$ is a misnomer;
one could {\it in extremis} `quantize' the coordinates of $V_0$.  The
observables will be some subset of the hermitian elements of 
${\cal A}_\kbar$.  We shall not discuss this problem here; we shall
implicitly suppose that all hermitian elements of ${\cal A}_\kbar$ are
observables, including the `coordinates'.  We shall not however have
occasion to use explicitly this fact. If there is a gravitational field
then there must be some source, of characteristic mass $\mu$. If 
$\mu^2 \kbar$ tends to zero with $\kbar$ then $V_0$ will be without
curvature.  We are interested here in the case in which $\mu^2 \kbar$
tends to some finite non-vanishing value as $\kbar \rightarrow 0$.

Quite generally the commutator of ${\cal A}_\kbar$ defines a Poisson
structure (Vaisman 1994) on $V_0$.  We have given arguments in the past
(Madore \& Mourad 1996b), based on simple models, that a differential
calculus over ${\cal A}_\kbar$ should determine a metric-compatible
torsion-free linear connection on $V_0$. It is natural then that there
should be a relation between the Poisson structure and the curvature of
the connection. We can show that certain natural hypotheses on the
differential calculus yield in fact relations between the two.  We have
been however unable so far to present a realistic gravitational field 
explicitly as the `shadow' of a differential calculus over a
noncommutative algebra.

In the next section we make some general remarks concerning the problem
of `quantization' of space-time.  In Section~3 we discuss in particular
the case of a dynamical space-time which is asymptotically flat.  Greek
indices take values from 0 to 3; the first half of the alphabet is used
to index frames and the second half to index generators.  Latin indices
take values from 0 to $n-1$.  The bracket $[\xi, \eta]$ of two elements
$\xi$ and $\eta$ of a graded algebra is a graded bracket.

\section{The general formalism}

We define 6 more elements $q^{\mu\nu}$ of ${\cal A}_\kbar$ by the
relations
$$
[q^\mu, q^\nu] = i  \kbar q^{\mu\nu}.                              \eqno(2.1)
$$
The details of the structure of ${\cal A}_\kbar$ will be contained, for
example, in the commutation relations $[q^\lambda, q^{\mu\nu}]$.  The
$q^{\mu\nu}$ can be also considered as extra generators and the
Equations~(2.1) as extra relations. In this case the $q^{\mu\nu}$
cannot be chosen arbitrarily. They must satisfy the four Jacobi identities:
$$
[q^\lambda, q^{\mu\nu}] + [q^\mu, q^{\nu\lambda}] + 
[q^\nu, q^{\lambda\mu}] = 0.                                       \eqno(2.2) 
$$
We define recursively an infinite sequence of elements by setting,
for $p \geq 1$,
$$
[q^\lambda, q^{\mu_1 \cdots \mu_p}] = 
i \kbar q^{\mu_1 \cdots \mu_{(p+1)}}.                              \eqno(2.3)
$$
Several structures have been considered in the past, by Snyder (1947),
Madore (1989a, 1995) and Doplicher {\it et al.} (1994). With our choice
of normalization $q^{\mu_1 \cdots \mu_p}$ has units of mass to the
power $p-2$.

We shall assume that for the description of a generic gravitational
field the appropriate algebra ${\cal A}_\kbar$ has a trivial center
${\cal Z}_\kbar$:
$$
{\cal Z}_\kbar = {\mathbb C}.                                      \eqno(2.4) 
$$
The only argument we have in favour of this assumption is the fact that
it could be argued that if ${\cal Z}_\kbar$ is greater then 
${\mathbb C}$ then the `quantization' has been only partial. It implies
of course that the sequence of $q^{\mu_1 \cdots \mu_p}$ never ends,
although all these elements need not be independent.

The simplest example of an algebra is the tensor algebra ${\cal A}_u$
over the vector space spanned by the 4 generators $q^\mu$.  This algebra
has a natural filtration ${\cal F}^p_u$ in powers of $\kbar$.  One
defines ${\cal F}^0_u$ as the symmetric algebra over the generators. We
can identify this with the algebra of polynomials on the classical
space-time $V$ and, by completion, with the algebra of smooth functions.
One defines ${\cal F}^1_u$ as the kernel of the projection of ${\cal
A}_u$ onto ${\cal F}^0_u$. It is an ideal of ${\cal A}_u$ each of whose
elements contains at least one commutator.  One defines ${\cal F}^p_u$
as the ideal of those elements which contain at least $p$ commutators.
The most general algebra ${\cal A}_\kbar$ is a quotient of ${\cal A}_u$
by some ideal ${\cal I}$ and the filtration of ${\cal A}_u$ defines a
corresponding filtration ${\cal F}^p_\kbar$ of ${\cal A}_\kbar$.  There
is an intimate connection between the dimension of $V_0$ and the `size'
of ${\cal I}$. We set
$$
x^{\mu_1 \cdots \mu_p} = 
\lim_{\kbar \rightarrow 0}q^{\mu_1 \cdots \mu_p}.
$$
A set of independent elements of the complete set of the 
$x^{\mu_1 \cdots \mu_p}$ are local coordinates of $V_0$. If $V$ is
Minkowski space-time then the condition of Lorentz invariance in the
commutative limit forces $x^\lambda$ and at least 4 of the 6 coordinates
$x^{\mu\nu}$ to be independent (Doplicher {\it et al.} 1994).  In
general the $x^{\mu_1 \cdots \mu_p}$ for $p \geq 3$ can at least in part
be functions of $x^\lambda$ and $x^{\mu\nu}$.  This will depend on the
structure of the ideal ${\cal I}$. If ${\cal I} = 0$ then all the
$x^{\mu_1 \cdots \mu_p}$ for $p \geq 3$ are independent coordinates and
$\mbox{dim} V_0 = \infty$. If on the other hand 
$x^{\mu_1 \cdots \mu_p} = x^{\mu_1 \cdots \mu_p}(x^\lambda)$ for all 
$p \geq 2$ then $V_0 = V$.

Let $f$ and $g$ be elements of ${\cal A}_\kbar$. To simplify the
notation we use the same symbol to designate the corresponding limit
functions on $V_0$. The Poisson bracket is given by
$$
\{f,g\} = \lim_{\kbar \rightarrow 0} {1\over i\kbar} [f,g].         \eqno(2.5)
$$
The restriction of the bracket to ${\cal A}$ is given by
$$
\{f,g\} = x^{\mu\nu} \partial_\mu f \partial_\nu g.                 \eqno(2.6On)
$$

Let $\Omega^*({\cal A}_\kbar)$ be a differential calculus over 
${\cal A}_\kbar$.  We shall assume that $\Omega^*({\cal A}_\kbar)$ is
non-degenerate in the sense that if $df = 0$ then 
$f \in {\cal Z}_\kbar$.  This is a natural assumption.  It is easy to
see that $\Omega^1({\cal A}_\kbar)$ is generated by the $dq^\lambda$ as
a bimodule.  It is typical in fact of noncommutative differential
calculi that $\Omega^1({\cal A}_\kbar)$ have one generator as a
bimodule and that the $dq^\lambda$ can be written in turn as a
commutator of a sort of generalized `Dirac operator' (Connes 1986).  We
shall suppose that $\Omega^1({\cal A}_\kbar)$ is a free left (and right)
${\cal A}_\kbar$-module of rank $n$ with a basis $\theta^a$, 
$0 \leq a \leq n-1$, which commutes with the generators $q^\lambda$ of
${\cal A}_\kbar$:
$$
[q^\lambda, \theta^a] = 0.                                          \eqno(2.7)
$$ 
In view of the fact that we suppose that our algebras are
`quantizations' of parallelizable manifolds this is a natural
assumption.  We consider $\theta^a$ as the noncommutative analogue of a
moving frame. We use the word `frame' or `Stehbein' since, as there are
no points, there can be no `movement'.  It is easy to see that $n$ must
be bounded below by the dimension of $V_0$:
$$
n \geq  \mbox{dim} V_0.                                            \eqno(2.8)
$$

A rather trivial example of a differential calculus is the universal
differential calculus which exists over every algebra. Its structure as
a graded algebra is rather trivial and it is comforting to know that it
admits naturally a trivial torsion-free linear connection (Madore \&
Mourad 1996b).  If we consider only differential calculi which are based
on sets of derivations (See, for example, Dimakis \& Madore 1996) then
the dual basis will automatically satisfy the condition (2.7). Of course
any moving frame on $V_0$ satisfies an analogous condition.  If
$\theta^\alpha = dx^\alpha$ on an ordinary manifold then the associated
torsion-free linear connection is flat.  The noncommutative equivalent
(2.7) would imply that $dq^{\mu_1 \cdots \mu_p} = 0$ for all $p \geq 2$.
Because of our assumptions this means that $q^{\mu_1 \cdots \mu_p} = 0$
for all $p \geq 2$ and that the algebra is commutative.

An interesting question is the relation between the limit
$$
\Omega^*_0 = \lim_{\kbar \rightarrow 0} \Omega^*({\cal A}_\kbar)
$$
of a differential calculus over ${\cal A}_\kbar$ and the de~Rham 
differential calculus $\Omega^*(V_0)$ over $V_0$. The universal calculus 
$\Omega^*_u({\cal A}_\kbar)$ obviously does not have $\Omega^*(V_0)$ as
a limit. It would seem that (2.7) is a necessary condition for this to
be true if $V_0$ is parallelizable. It is not obvious that the dependence
on a set of derivations is a sufficient condition. Suppose that 
$\Omega^*_0 = \Omega^*(V_0)$.  The commutator on 
$\Omega^*({\cal A}_\kbar)$ induces, in the commutative limit, a Poisson
bracket on $\Omega^*(V_0)$. Let $\xi$ and $\eta$ be two elements of
$\Omega^*({\cal A}_\kbar)$ which to be specific we shall suppose are
1-forms. We use the same symbol to designate the corresponding de~Rham
1-forms. This notation can be misleading since it can happen (Madore
1995) that as an element of $\Omega^1({\cal A}_\kbar)$ $\xi$ is exact
but that as element of $\Omega^1(V_0)$ it is only closed. The Poisson
bracket of $\xi$ and $\eta$ is given by the same formula (2.5) as for
the functions.  Obviously the Jacobi identity remains satisfied but the
bracket of two 1-forms need not be an element of $\Omega^*(V_0)$. To see
this we write $\xi = \xi_a \theta^a$ and $\eta = \eta_a \theta^a$. Then
$$
[\xi, \eta] = [\xi_a, \eta_b] \theta^a \theta^b + 
\eta_b \xi_a [\theta^a, \theta^b].
$$
The first term on the right-hand side tends obviously to the element
$\{\xi_a,\eta_b\} \theta^a \theta^b$ of $\Omega^2(V_0)$. The limit of
the second term however will be in general a more general element of
$\Omega^1(V_0) \otimes \Omega^1(V_0)$. There is an interesting case in
which a `twisted' bracket $\{\xi, \eta\}_C$ can be defined (Chu {\it 
et al.} 1996) which takes its values in $\Omega^*(V_0)$. We suppose that
the $\theta^a$ satisfy relations of the form (Dimakis \& Madore 1996)
$$
\theta^a \theta^b + C^{ab}{}_{cd} \theta^c \theta^d = 0
$$
where the $C^{ab}{}_{cd}$ are elements of the center of the algebra such
that 
$$
\lim_{\kbar \rightarrow 0} C^{ab}{}_{cd} = \delta^b_c \delta^a_d. 
$$
Then we can define
$$
\{\xi, \eta\}_C = \lim_{\kbar \rightarrow 0} {1\over i\kbar}
(\xi_a \eta_b - C^{cd}{}_{ab} \xi_c \eta_d) \theta^a \theta^b.      \eqno(2.9)
$$

There always exist elements 
$\theta^{\mu_1 \cdots \mu_p}_a \in {\cal A}_\kbar$ such that
$$
dq^{\mu_1 \cdots \mu_p} = \theta^{\mu_1 \cdots \mu_p}_a \theta^a.
$$
We have remarked that $\Omega^1({\cal A}_\kbar)$ is generated by the 4
elements $dq^\lambda$ as an ${\cal A}_\kbar$-bimodule. We shall suppose
that there exists a subset $dq^i$ of the $dq^{\mu_1 \cdots \mu_p}$,
with $0 \leq i \leq n-1$, which are also generators of 
$\Omega^1({\cal A}_\kbar)$ as a (left or right) ${\cal A}_\kbar$-module.
This means that there exist elements $\theta^a_i$ and 
$\theta^{\prime a}_i$ in ${\cal A}_\kbar$ such that one can write
$$
\theta^a = \theta^a_i dq^i = dq^i \theta^{\prime a}_i.
$$
It follows then that 
$$
dq^i f = \theta^i_a \theta^a f = \theta^i_a f \theta^a =
(\theta^i_a f \theta^a_j) dq^j.
$$
An element $f$ on the right of $dq^i$ can always be shifted to the left.
An explicit counter-example of this is given in Section~IV of the
article by Dimakis \& Madore (1996).

As an example choose $n = 10$ and set
$\theta^a = (\theta^\alpha, \theta^{\beta\gamma})$ with the second
component antisymmetric in its two indices.  We have then the expansion
$$
\theta^\alpha = \theta^\alpha_\lambda dq^\lambda + 
{1\ \over 2} \theta^\alpha_{\mu\nu} dq^{\mu\nu}, \qquad
\theta^{\beta\gamma} = \theta^{\beta\gamma}_\lambda dq^\lambda + 
{1\ \over 2} \theta^{\beta\gamma}_{\mu\nu} dq^{\mu\nu}.
$$
We derive therefore from (2.7) the relations
$$
\theta^\alpha_\nu [q^\mu, dq^\nu] + 
[q^\mu, \theta^\alpha_\nu] dq^\nu +
{1\over 2}\theta^\alpha_{\rho\sigma} [q^\mu, dq^{\rho\sigma}] +
{1\over 2}[q^\mu, \theta^\alpha_{\rho\sigma}] dq^{\rho\sigma} = 0
                                                                   \eqno(2.12)
$$
and
$$
\theta^{\beta\gamma}_\nu [q^\mu, dq^\nu] + 
[q^\mu, \theta^{\beta\gamma}_\nu] dq^\nu +
{1\over 2}\theta^{\beta\gamma}_{\rho\sigma} [q^\mu, dq^{\rho\sigma}] +
{1\over 2}[q^\mu, \theta^{\beta\gamma}_{\rho\sigma}] dq^{\rho\sigma} = 0.
                                                                   \eqno(2.13)
$$
These equations can in principle be solved to yield expressions for the
commutators $[q^\mu, dq^\nu]$ and $[q^\lambda, dq^{\mu\nu}]$.

We have not succeeded in constructing a linear connection compatible
with this frame in the sense of Madore \& Mourad (1996b). We have chosen
$n = 10$ so that there will be the correct number of degrees of freedom
to yield a general metric but this is not of course a sufficient
condition. It must be shown that the frame is dual to a set of
derivations which satisfies compatibility conditions.  We cannot show
that the Riemann tensor we use below is is necessarily the limiting
curvature of a linear connection on a noncommutative structure.

\section{The asymptotically flat case}

Consider now a differential calculus $\Omega^*({\cal A}_\kbar)$ over an
algebra ${\cal A}_\kbar$ which is asymptotically flat in the sense that
in the commutative limit there exists a radial coordinate $r$ such that
$$
\theta^\alpha_\lambda = \delta^\alpha_\lambda + o(r^{-1}), \qquad
\theta^\alpha_{\mu\nu} = o(r^{-1}).                                 \eqno(3.1)
$$
Because of (2.8) the dimension of $V_0$ is at most equal to 10. Let
$(x^\lambda, x^{\mu\nu})$ be local coordinates and $f$ an arbitrary
element of ${\cal A}_\kbar$.  We can write then in the `quasi-classical'
approximation
$$
[q^\lambda, f] = i \kbar x^{\lambda\mu} \partial_\mu f +
{1 \over 2} i\kbar x^{\lambda\mu\nu} \partial_{\mu\nu} f + o(\kbar^2).
                                                                    \eqno(3.2)
$$
On the right-hand side one can consider $f$ as a function on $V_0$.
We deduce from (2.12) in the limit $\kbar \rightarrow 0$ that
$$
\eqalign{
dx^{\mu\nu} + 
x^{[\mu\rho} \partial_\rho \theta^{\nu]}_\sigma dx^\sigma &+
{1\over 2} x^{[\mu\rho\sigma} 
\partial_{\rho\sigma} \theta^{\nu]}_\tau dx^\tau +
{1\over 2}\theta^\alpha_{\rho\sigma} \{q^\mu, dq^{\rho\sigma}\} \cr &+
{1\over 2} x^{[\mu\lambda} 
\partial_\lambda \theta^{\nu]}_{\rho\sigma} dx^{\rho\sigma} + 
{1\over 2} x^{[\mu\rho\sigma} 
\partial_{\rho\sigma} \theta^{\nu]}_{\lambda\tau} 
dx^{\lambda\tau} = 0
}                                                                   \eqno(3.3)
$$
and we conclude that
$$
dx^{\mu\nu} = o(r^{-1}).                                            \eqno(3.4)
$$
Therefore the extra variables are dependent,
$$ 
x^{\mu\nu} = x^{\mu\nu}(x^\lambda)                                  \eqno(3.5)
$$
and the Poisson structure varies from point to point. We have $V_0 = V$
and $\mbox{dim} V_0 = 4$.  Using (3.4) we can conclude that the last 4
terms in (3.3) are of order $o(r^{-2})$ and we can rewrite (3.3) as
$$
dx^{\mu\nu} + 
x^{[\mu\rho} \partial_\rho \theta^{\nu]}_\sigma dx^\sigma =
o(r^{-2}).                                                          \eqno(3.6)
$$
This system of equations has integrability conditions which yield a
relation between the symplectic structure and the curvature.

One can be more explicit when the gravitational field is dynamical with
a smooth asymptotic expansion along a family of retarded null hypersurfaces.
We can write then the derivatives of the components $\theta^\alpha_\mu$
of the moving frame on $V$ to leading order in the asymptotic expansion as
$$
\partial_\lambda \theta^\alpha_\mu = 
p_\lambda \dot \theta^\alpha_\mu + o(r^{-2})
$$
with $p_\lambda$ a null vector and the dot derivation with respect to
the function which parameterizes the family of hypersurfaces. Then
Equation~(3.6) becomes
$$
\dot x^{\mu\nu} p_\sigma + 
x^{[\mu\rho} p_\rho \dot \theta^{\nu]}_\sigma = o(r^{-2}).          \eqno(3.7)
$$

Define $\tilde p^\mu = x^{\mu\nu} p_\nu$. Equation (3.7) has the
particular solution 
$$
\dot x^{\mu\nu} = o(r^{-2}), \qquad
\tilde p_{\phantom{\rho}}^{[\mu} \dot \theta^{\nu]}_\rho = o(r^{-2})
$$ 
which yields the relation
$$
\tilde p_{[\mu} \ddot g_{\nu][\rho} \tilde p_{\sigma]} = o(r^{-2}). \eqno(3.8)
$$ 
On the other hand the asymptotic expression for the Riemann tensor is 
given by
$$
R_{\mu\nu\rho\sigma} =
p_{[\mu} \ddot g_{\nu][\rho} p_{\sigma]} + o(r^{-2}).               \eqno(3.9)
$$
To leading order the Riemann tensor is of null type with $p_\mu$ as
principle vector.  We see then from this simple example that there is an
intimate relation between the Riemann tensor and the Poisson structure
of space-time; we can write
$$
[q^\mu, q^\nu] = i \kbar x^{\mu\nu} + o(\kbar^2)                   \eqno(3.10)
$$
where the right-hand side is strongly restricted if not determined by
the Riemann tensor.

\section{Discussion}

We have found, to lowest order in the `quantization' parameter an
asymptotic form of the commutation relations which depends on the form
of the asymptotic curvature.  We are free to suppose that the
commutation relations vanish at infinity but it would be quite
consistent to suppose that they tend to constant values. An algebra has
been proposed (Doplicher {\it et al,} 1994) for which the fundamental
commutator of the generators is constant everywhere and yet possesses
(Madore \& Mourad 1996b) a differential calculus with a torsion-free
metric compatible flat linear connection. The most important assumption
we have made concerning the differential calculus is contained in 
Equation~(2.7).

\section*{Acknowledgments} 

Part of this research was done while the author was visiting the
Ludwig-Maximilians-Universit\"at, M\"unchen. He would like to thank 
J.~Wess for his hospitality and interesting conversations.

\parindent=0cm
\tolerance=1000
\parskip 4pt plus 1pt

\section*{References}

Chu Chong-Sun, Ho Pei-Ming, Zumino B. 1996, {\it Some Complex Quantum
Manaifolds and their Geometry}, Lectures given at the NATO Advanced Study
Institute on Quantum Fields and Quantum Space Time, Cargese, hep-th/9608188.

Connes A. 1986, {\it Non-Commutative Differential Geometry}, Publications
of the Inst. des Hautes Etudes Scientifique. {\bf 62} 257.

Dimakis A., Madore J. 1996, {\it Differential Calculi and Linear 
Connections}, J. Math. Phys. {\bf 37} 4647.

Doplicher S., Fredenhagen K., Roberts, J.E. 1994, 
{\it Spacetime quantization induced by classical gravity}, 
Phys. Lett. {\bf B331} 39.

Dubois-Violette M., Kerner R., Madore J.  1997, 
{\it Shadow of Noncommutativity}, 
Preprint LPTHE Orsay 96/06, q-alg/9702030.

Gerstenhaber M. 1964, {\it On the deformation of rings and algebras},
Ann. Math. {\bf 79} 59.

Madore J. 1989a, {\it Non-Commutative Geometry and the Spinning Particle}, 
XI Warsaw Symposium on Elementary Particle Physics, May, 1988, Kazimierz.

--- 1989b, {\it Kaluza-Klein Aspects of Noncommutative Geometry}, 
Proceedings of the XVII International Conference on Differential
Geometric Methods in Theoretical Physics, August, 1988, Chester.

--- 1995, {\it An Introduction to Noncommutative Differential Geometry
and its Physical Applications}, Cambridge University Press.

--- 1997, {\it Fuzzy Space-time}, Can. J. Phys. (to appear), gr-qc/9607065.

Madore J., Mourad. J. 1996a, 
{\it Noncommutative Kaluza-Klein Theory}
Lecture given at the 5th Hellenic School and Workshops on Elementary
Particle Physics, hep-th/9601169.

Madore J., Mourad. J. 1996b, 
{\it Quantum Space-Time and Classical Gravity}, 
Preprint, LPTHE Orsay 95/56, gr-qc/9607060.

Snyder H.S. 1947, {\it Quantized Space-Time}, Phys. Rev. {\bf 71} 38.

Vaisman Izu 1994, {\it Lectures on the Geometry of Poisson Manifolds},
Birkh\"auser Verlag, Basel.

\end{document}